# Spectral Variability of the Star HD 141569A

H. N. Adigozalzade[1], N. Z. Ismailov[1], U. Z. Bashirova, and S. A. Alishov
Tusi Shamakhy Astrophysical Observatory, Ministry of Science and Education, Shamakhy AZ5626, Azerbaijan;
hadigozalzade@gmail.com

## Abstract

The paper presents the results of long-term, homogeneous spectroscopic studies of the Herbig Ae/Be star HD 141569A, obtained in the optical range. We detect spectral variability on timescales of months to years, including changes in the spectrophotometric parameters of the H$\alpha$ emission line, as well as other hydrogen absorption lines. Based on the derived physical parameters and age, the star appears to be in a transitional evolutionary stage—from a protoplanetary disk to a debris disk. We determine that the star's axial rotation period is approximately 12 hr, and its equatorial velocity is significantly below the critical rotation velocity. The observed spectral variability is likely caused by partial obscuration of the central star's radiation by extended fragments of the disrupted disk.

## 1. Introduction

According to numerous studies, the star HD 141569A has a spectral classification of B9–A1 V (see, for example, Murphy et al. 2020; Vieira et al. 2003). It is considered a transitional object between young Herbig Ae/Be (HAeBe) stars and more evolved Vega-like or $\beta$ Pic-type stars (Malfait et al. 1998). HD 141569A is the primary component of a visual triple system, accompanied by two late-type companions (B and C), which are separated by approximately 8″ (Weinberger et al. 2000). These companions share the same proper motion and distance as the primary. The circumstellar disk of HD 141569A has been directly imaged using coronagraphic techniques (Augereau et al. 1999; Weinberger et al. 1999). Optical and near-infrared (near-IR) observations reveal a complex ring-like structure in the disk that extends to a radial distance of about 600 au. HD 141569A exhibits many characteristics typical of dusty disks associated with young main-sequence (MS) stars. However, the gas content of its disk is significantly higher than that of classical debris disks, as indicated by millimeter-wave and near-IR CO observations (Zuckerman et al. 1995; Boccaletti et al. 2003). According to Folsom et al. (2012), the effective temperature of the star and the surface gravity index are $T_{\rm eff} = 9800 \pm 500$ K and $\log g = 4.2 \pm 0.4$, respectively. Fairlamb et al. (2015) derived similar values: $T_{\rm eff} = 9750 \pm 250$ K, $\log g = 4.35 \pm 0.15$. Table 1 summarizes the star's main physical parameters, compiled from various scientific sources. According to observations by Mendigutía et al. (2011), H$\alpha$ line emission has an equivalent width (EW) of EW = $-6.0 \pm 0.1$ Å, and the width of the wings at 10% of peak intensity (W10 H$\alpha$) is approximately 641 km s$^{-1}$. The EWs of the [O I] 6300 Å, He I 5876 Å, and Na I D lines are also reported. The authors did not detect significant variability in the H$\alpha$ line for timescales from days to months.

Acke et al. (2005) measured an H$\alpha$ EW of –6.7 Å and a full width at half maximum (FWHM) of approximately 154 km s$^{-1}$ for the [O I] 6300 line. Lecavelier des Etangs et al. (2005) reported that the star did not exhibit significant brightness variations. The radial velocity (RV) measurements of the star vary between +3 km s$^{-1}$ and –7.5 km s$^{-1}$ (see, for example, Matthias et al. (2020); Gaia Collaboration (2022)).

In the initial studies, the star was not included in the lists of classic works on HAeBe stars (see, for example, Herbig 1960; Finkenzeller & Mundt 1984). Only after studying its infrared (IR) spectrum was it accepted as an HAe (Finkenzeller & Jankovics 1984) Be star. Spectroscopic studies of HD 141569A in the optical range were carried out sporadically, in individual observations (see, for example, Alecian et al. 2013; Fairlamb et al. 2015) within the framework of survey studies of program objects. In such observations, it is impossible to perform rapid or seasonal variation studies of the star's spectrum. The goal of our work is to investigate variations in the star's spectrum based on long-term, homogeneous spectroscopic observations.

## 2. Observations

Spectral observations of the star HD 141569A were carried out at the Cassegrain focus of the 2 m telescope at the Shamakhy Astrophysical Observatory, named after N. Tusi, using the ShaFES echelle spectrograph. The CCD camera STA4150A, with 4K × 4K elements and cooled with liquid nitrogen, was used as the light detector. The pixel size is 15 × 15 $\mu$m, and the working spectral range covers λ 3700–7800 Å. The spectral resolution at 2 × 2 binning was R ≈ 28,000 near the H$\alpha$ line. Spectral processing was performed using the current version of DECH software (http://www.gazinur.com/DECH-software.html). In addition to the spectra of the studied object and the standard star, we obtained a set of calibration frames: dark, flat field, and ThAr or Sky. The flat field was used for both per-pixel sensitivity correction and removing interference fringes. The spectrum of solar radiation scattered in the atmosphere was used for wavelength calibration, which was simultaneously tested by the spectrum of a ThAr lamp. This procedure allows one to obtain the positions of the reference lines in the echelle dispersion curve with an accuracy of no worse than ±0.01.

---
[1] Corresponding authors, these authors contributed equally to this work.

**Table 1**
Physical Parameters of the HD 141569A Collected from Literature

| D (pc) | $L/L_\odot$ | $M/M_\odot$ | $R/R_\odot$ | t (Myr) | vsini (km s$^{-1}$) | RV (km s$^{-1}$) | References |
|---|---|---|---|---|---|---|---|
| 116 ± 9 | 1.49 ± 0.06 | 2.33 ± 0.2 | 1.94 ± 0.21 | 5.7 ± 1.3 | 228 ± 10 | −12 ± 7 | Alecian et al. (2013), van Leeuwen (2007) |
| 99 | … | 2.2 | … | … | … | … | Mendigutia et al. (2011) |
| 111.13 | 1.4 | 2.12 | 1.75 | 7.97 | … | … | Guzman- Deaz et al. (2021) |
| 99 | 1.1 | … | … | … | … | −2 | Acke et al. (2005) |
| 112 ± 34 | 1.28 ± 0.3 | 1.9 ± 0.4 | 1.5 ± 0.5 | 9 ± 4.5 | … | … | Fairlamb et al. (2015) |
| … | … | … | … | … | 258 ± 17 | … | Andrillat et al. (1990) |
| 108 ± 6 | 1.411 | 2.00 ± 0.01 | 1.7 | 4.71 ± 0.3 | 236 ± 15 | … | Merín et al. (2004) |
| 207 | … | … | … | … | 236 ± 9 | −6 ± 5 | Dunkin et al. (1997) |

**Table 2**
Data on Spectral Observations of the Star HD 141569A

| Date | JD 2450000+ | t (s) | S/N |
|---|---|---|---|
| 20.05.2017 | 7894.3710 | 3000 | 90 |
| 06.06.2017 | 7911.2305 | 2400 | 100 |
| 02.05.2018 | 8240.5472 | 2400 | 90 |
| 22.05.2018 | 8260.5833 | 2400 | 150 |
| 25.05.2018 | 8264.2451 | 2400 | 120 |
| 06.06.2018 | 8276.2417 | 2400 | 100 |
| 07.06.2018 | 8277.2256 | 2400 | 110 |
| 27.06.2018 | 8297.2312 | 2400 | 120 |
| 08.05.2019 | 8612.3034 | 3600 | 120 |
| 31.05.2019 | 8635.4111 | 3000 | 110 |
| 01.06.2019 | 8636.2972 | 3600 | 100 |
| 02.06.2019 | 8637.4284 | 3600 | 90 |
| 05.06.2019 | 8640.2604 | 3000 | 110 |
| 13.06.2019 | 8648.2555 | 3600 | 120 |
| 23.06.2019 | 8658.2562 | 3000 | 120 |
| 24.06.2019 | 8659.2500 | 2700 | 115 |
| 03.07.2019 | 8668.2423 | 2400 | 95 |
| 08.07.2019 | 8673.2375 | 2400 | 85 |
| 03.06.2020 | 9004.3937 | 2400 | 110 |
| 12.06.2020 | 9013.2826 | 3000 | 120 |
| 03.07.2020 | 9034.2289 | 2400 | 110 |
| 08.07.2022 | 9769.2548 | 1800 | 120 |
| 18.07.2022 | 9779.2416 | 1800 | 90 |
| 06.07.2023 | 10132.3013 | 3000 | 100 |

A detailed description of the spectrograph can be found in Mikailov et al. (2020), while results from its application are presented in Ismailov et al. (2024) and Adigozalzade et al. (2025b). Table 2 provides the observation log for the spectral data collected. Observations were conducted between 2017 and 2023, during which a total of 24 pairs of spectrograms (target and standard star) were obtained. These were averaged over each observation night. The average signal-to-noise ratio (S/N) near the H$\alpha$ line was approximately 100; in the blue region of the spectrum, the S/N was approximately half as much. EWs and RVs were measured for hydrogen lines H$\alpha$–H$\varepsilon$, He I 5876, Na I D$_1$, D$_2$, and the forbidden line [O I] 6300 Å. The average measurement error for EWs of hydrogen lines, derived from standard star spectra, was up to 4%. For weaker lines, the error was around 6%. The uncertainties of the position measurement did not exceed ±1.5 km s$^{-1}$, and the intensity determinations were accurate to approximately 5%.

## 3. H$\alpha$ Line

Figure 1 presents the average profile of the H$\alpha$ line, obtained by averaging the normalized profiles of all spectra. Heliocentric RVs are plotted along the abscissa. In a separate panel of Figure 1, the standard deviation of the mean intensity across the wavelength range of the profile is shown below. As can be seen, the H$\alpha$ line exhibits two emission peaks, with an intensity ratio $I_V/I_R > 1$, consistently observed across all H$\alpha$ profiles. A similar stable intensity ratio of emission components in the H$\alpha$ line had been observed also by other authors (see, e.g., Mendigutía et al. 2011; Alecian et al. 2013;

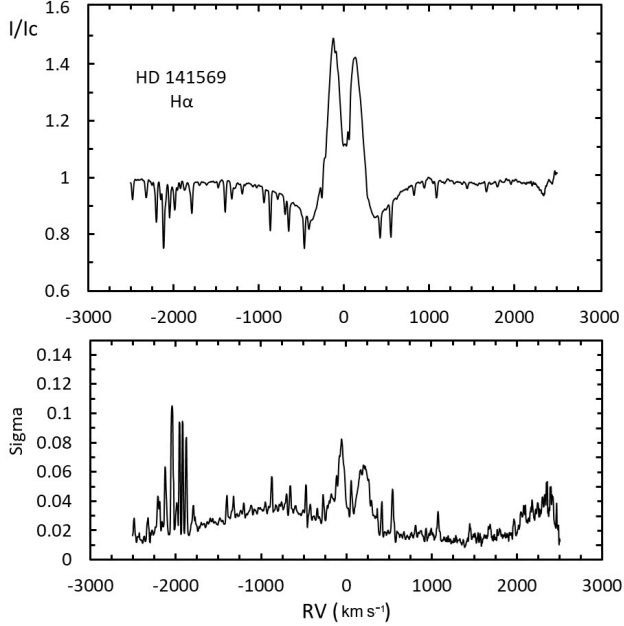

**Figure 1.** The average profile of the H$\alpha$ line obtained over 6 yr (upper panel) and the profile of its standard deviation ($\sigma$) from the average (lower panel). The level of measurement errors for sigma is about 5%.

Ababakr et al. 2017). As can be seen in Table 3, the average EW of the absorption component of H$\alpha$ is EWa1 = 1.98 ± 0.67 Å, and for the red component, EWa2 = 1.39 ± 0.43 Å. A similar asymmetric profile (of the P Cyg type) was also observed in the work of Mendigutía et al. (2011). Since the direction of the star's rotation axis is approximately 57° ± 1° (Di Folco et al. 2020) to the line of sight, we assume that this intensity ratio of the H$\alpha$ line emission components is related to the orientation of the star's disk in space.

The average RVs of the individual peaks and the central absorption trough are $-126.6 \pm 5.9$ km s$^{-1}$, $120.9 \pm 9.3$ km s$^{-1}$, and $7.1 \pm 15.8$ km s$^{-1}$, respectively. The bisector velocity of the emission is $-0.4 \pm 4.0$ km s$^{-1}$, indicating that the emission center shift (bisector velocity) closely matches the systemic velocity of the star. All profile components exhibit RV variations that are slightly greater than the measurement uncertainties. However, no significant variations in the overall H$\alpha$ profile have been observed on timescales from days to months. From the lower panel of Figure 1, it is evident that the most pronounced variations occur in the intensities of the emission peaks. The maximum deviation from the mean at the peak positions reaches 7%–8%, which is slightly higher than the associated measurement errors.

Table 3 presents the results of the measured parameters of the H$\alpha$ spectral line. From left to right, the columns show the EWs of the blue EWa1 and red EWa2 absorption and the emission peak EWe, the emission full-width at half maximum FWHM, the RVs of the blue and red emission peaks RV1p and RV2p, the bisector emission velocity RVbis, and the central absorption velocity RVea. The last two lines of Table 3 present the average values of the parameters and their corresponding standard deviations. Here, too, it is evident that the greatest change in the RVs is observed at the peak tops and at the central absorption.

Figure 2 shows the time variation of the EWs of the H$\alpha$ line components: emission (EWe), blue absorption (EWa1), and red absorption (EWa2). A gradual increase in emission from 2017 to 2019, followed by a decline in subsequent years, is accompanied by an inverse trend in the blue and red absorption components (EWa1 and EWa2). The EWs of these profile components vary on a timescale of approximately one year.

Figure 3 displays a similar graph to Figure 2, but is limited to the data obtained in 2019. This format allowed us to trace changes in the EWs of emission and absorption components of the H$\alpha$ line, as well as the RVs of the profile components, throughout the year. As shown in Figure 3, the EWs of the absorption components vary in opposition to those of the emission component, with smooth transitions. The entire observation season spans only 70 days. Toward the end of the 2019 season, an increase in emission is observed, along with a corresponding decrease in the EWs of the H$\alpha$ line's absorption components. Meanwhile, the RV components of the emission remain almost unchanged. Figure 4 shows the time variability of the RVs of the individual emission peaks in the H$\alpha$ line, as well as the bisector RV of the emission–defined as the velocity of the line center relative to the wings at half the peak intensity. As also shown in Table 3, the greatest deviations from the mean are observed for the red emission peak (RVp2 = $120.9 \pm 9.3$ km s$^{-1}$) and the central dip between the peaks (RVea = $7.1 \pm 15.8$ km s$^{-1}$).

The blue emission peak shows a shift of $-126 \pm 5.9$ km s$^{-1}$. In contrast, the bisector velocity (RVbis) varies only slightly ($-0.4 \pm 4.0$ km s$^{-1}$). The phrase can be close to the systemic velocity of the star (see Table 1). A comparison of these parameters with the data from Merín et al. (2004) shows satisfactory agreement. Analysis of the observational data reveals a high degree of correlation between the EWs of the emission and absorption components. Figure 5 shows the graphs between the parameters EWe and EWa1 (inverse correlation, $r = -0.09 \pm 0.09$), between EWa2 and EWa1 (direct correlation, $r = 0.68 \pm 0.15$), between RVp1 and RVp2 (direct, $r = 0.75 \pm 0.12$), and between RVbis and RVp1 ($r = 0.62 \pm 0.16$). The correlation coefficient $r$ between the EWe and EWa1 parameters is insignificant. The average RV of the absorption component in the blue wing is approximately $-350$ km s$^{-1}$ (see, Figure 1), which significantly exceeds the RV of the peak of the blue emission component. The lack of correlation between these parameters indicated the existence of different mechanisms for their variability: the absorption component originates in the stellar wind, while the emission peaks form in the circumstellar disk.

**Table 3**
Spectrophotometric Parameters of the Hα Line

| JD 245,0000+ | EWa1 (Å) | EWe (Å) | EWa2 (Å) | FWHM (Å) | RV1p (km s$^{-1}$) | RV2p (km s$^{-1}$) | RVbis (km s$^{-1}$) | RVea (km s$^{-1}$) |
|---|---|---|---|---|---|---|---|---|
| 7894.3710 | 2.01 | 2.64 | 1.30 | 8.21 | −130 | 118 | −3.09 | 1.61 |
| 7911.2305 | 2.60 | 1.85 | 1.93 | 8.15 | −134 | 122 | 4.91 | −2.11 |
| 8240.5472 | 2.04 | 2.35 | 1.38 | 8.27 | −124 | 123 | −0.76 | −11.98 |
| 8260.5833 | 2.58 | 2.45 | 1.59 | 8.08 | −129 | 119 | 0.54 | −7.38 |
| 8264.2451 | 1.91 | 2.70 | 1.08 | 8.13 | −137 | 107 | −4.03 | −15.70 |
| 8276.2417 | 2.13 | 2.88 | 0.98 | 8.30 | −130 | 112 | −1.58 | −1.58 |
| 8277.2256 | 1.78 | 3.04 | 1.01 | 8.61 | −125 | 133 | 3.00 | 8.61 |
| 8297.2312 | 3.23 | 2.05 | 1.39 | 8.34 | −122 | 135 | 0.32 | 3.37 |
| 8612.3034 | 1.65 | 2.95 | 1.11 | 8.40 | −122 | 120 | 0.95 | −10.28 |
| 8635.4111 | 1.67 | 3.05 | 1.22 | 8.33 | −132 | 118 | −3.93 | −1.01 |
| 8636.2972 | 1.43 | 2.97 | 1.73 | 8.41 | −135 | 104 | −1.32 | −10.66 |
| 8637.4284 | 1.55 | 3.02 | 1.02 | 8.42 | −124 | 122 | 7.03 | −9.43 |
| 8640.2604 | 1.53 | 2.93 | 1.26 | 8.24 | −131 | 115 | −2.99 | 38.74 |
| 8648.2555 | 0.85 | 3.35 | 0.91 | 8.59 | −133 | 118 | 3.2 | 3.20 |
| 8658.2562 | 1.23 | 3.13 | 1.06 | 8.40 | −124 | 122 | −6.00 | 26.02 |
| 8659.2500 | 1.28 | 3.27 | 0.94 | 8.66 | −123 | 115 | 0.11 | 24.12 |
| 8668.2423 | 0.71 | 3.72 | 0.69 | 8.75 | −124 | 128 | 3.08 | 43.15 |
| 8673.2375 | 1.97 | 3.04 | 1.21 | 8.54 | −124 | 130 | 3.60 | 26.50 |
| 9004.3937 | 3.02 | 2.55 | 1.81 | 7.84 | −129 | 111 | −5.24 | 4.98 |
| 9013.2826 | 2.15 | 2.20 | 2.39 | 8.01 | −121 | 120 | 2.62 | 12.79 |
| 9034.2289 | 3.15 | 2.20 | 1.82 | 7.88 | −113 | 141 | −5.45 | 10.55 |
| 9769.2548 | 2.23 | 2.32 | 1.69 | 8.13 | −127 | 112 | −5.92 | 8.60 |
| 9779.2416 | 2.34 | 2.06 | 2.07 | 8.07 | −130 | 119 | −5.36 | 15.09 |
| 10132.3013 | 2.59 | 2.03 | 1.78 | 7.66 | −115 | 137 | 6.21 | 13.44 |
| Mean | 1.98 | 2.70 | 1.39 | 8.27 | −126 | 120 | −0.42 | 7.11 |
| ±σ | 0.67 | 0.49 | 0.43 | 0.27 | 5.94 | 9.30 | 4.01 | 15.81 |

The detection of a high correlation between the EWs of the absorption components in the blue and red wings in Figure 1 indicates that, apparently, part of the mass ejected by the stellar wind is primarily returning to the star (see, Section 5). High-level positive correlation between the shifts of the emission component peaks indicated a common variability mechanism for RVs. This mechanism, for example, could be the kinematic passage of individual clumps in the circumstellar disk of the star. The bottom panel in the second part of Figure 1 shows the variation of the standard deviation of intensity σ of the Hα line profile as a function of wavelength. It is evident that the structure of the change in σ relative to zero velocity is asymmetric: the blue peak in the sigma plot is located at approximately 0 km s$^{-1}$, while the red component is clearly shifted by approximately +200 km s$^{-1}$. The arguments mentioned above indicate a significant role of the stellar wind in the variability of the star's spectrum.

In general, the results indicate that an increase in the EWs of the absorption components is accompanied by a decrease in emission intensity, and vice versa, stronger emission corresponds to weaker absorption lines. This behavior suggests that episodic stellar wind may replenish material into the circumstellar disk, enhancing emission features.

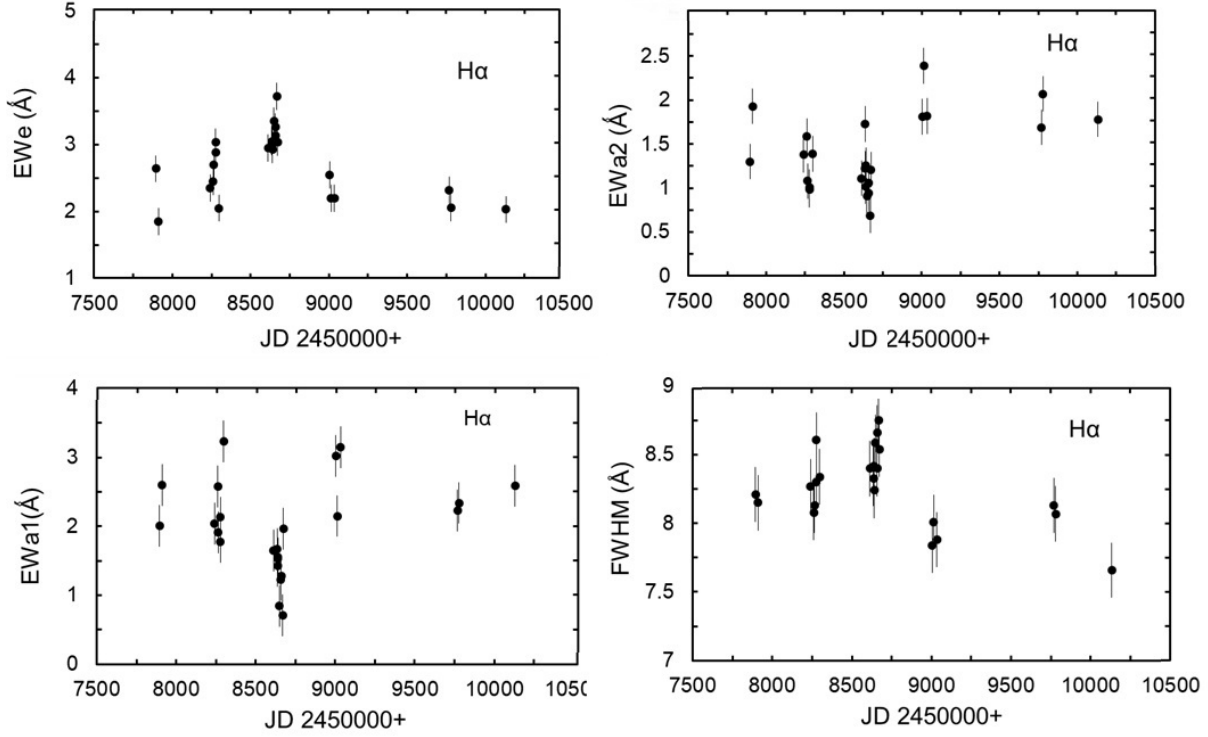

**Figure 2.** Changes in the EWs of the emission component (EWe), as well as the blue (EWa1) and red (EWa2) absorption components, and the half-width of the emission component FWHM of the H$\alpha$ line. The vertical bars show the average measurement error.

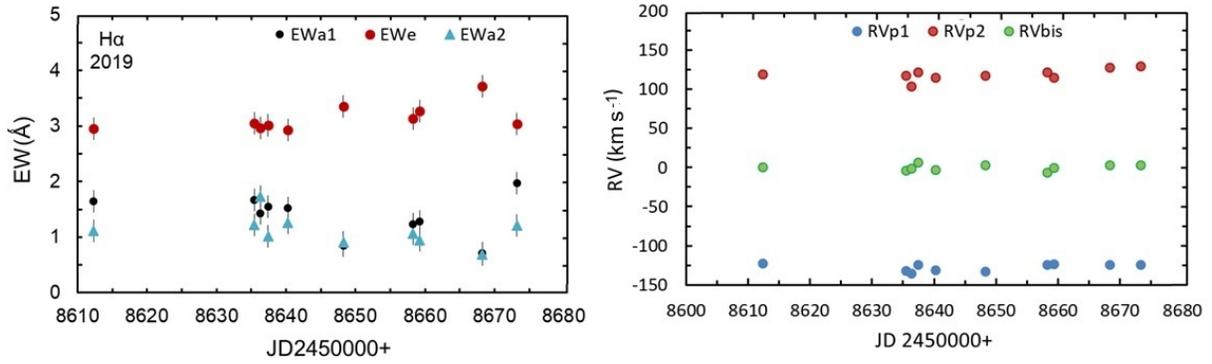

**Figure 3.** Temporal change of the H$\alpha$ line parameters for 2019. On the left panel, red dots show EWe, and black and blue dots show the parameters EWa1 and EWa2, respectively. On the right panel, RVs of the emission components in the H$\alpha$ line: blue dots RVp1, red dots RVp2, and bisector velocity (emission center) RVbis.

As can be seen from the hydrogen line profiles (Figures 1 and 6), the H$\alpha$ line exhibits a strong two-peak emission, while in the H$\beta$ line the emission is noticeable in a weak form only in the line wings. Emission is clearly not observed starting with the H$\gamma$ line and for subsequent Balmer series lines with a higher quantum number. This indicates a large steepness of the Balmer decrement in the disk, i.e., the stellar disk matter is optically thick in the hydrogen line emission. This may be partially caused by additional interstellar absorption in circumstellar space. The hydrogen line profiles in Figures 1 and 6 show only a small scatter from the average over the wings. We can only assume variability of intensity in the blue photospheric wing of the H$\alpha$ line (up to $2\% \pm 0.5\%$) (Figure 1), and simultaneously in the red wing of the H$\gamma$ line (Figure 6).

## 4. Other Spectral Lines

We also considered the change in the profiles of the absorption hydrogen lines H$\beta$ and H$\gamma$ for all the obtained spectra. Figure 6 displays the normalized profiles of the named hydrogen lines. The sigma value of the intensity is given at the bottom of each profile. As can be seen in Figure 6, the Hβ line features two weak emission peaks that are superimposed on the blue and red photosphere wings of the profile, respectively. The Hγ line does not show any clear signs of emission. According to the sigma level, it can be clearly stated that both lines did not show any clear signs of variability.

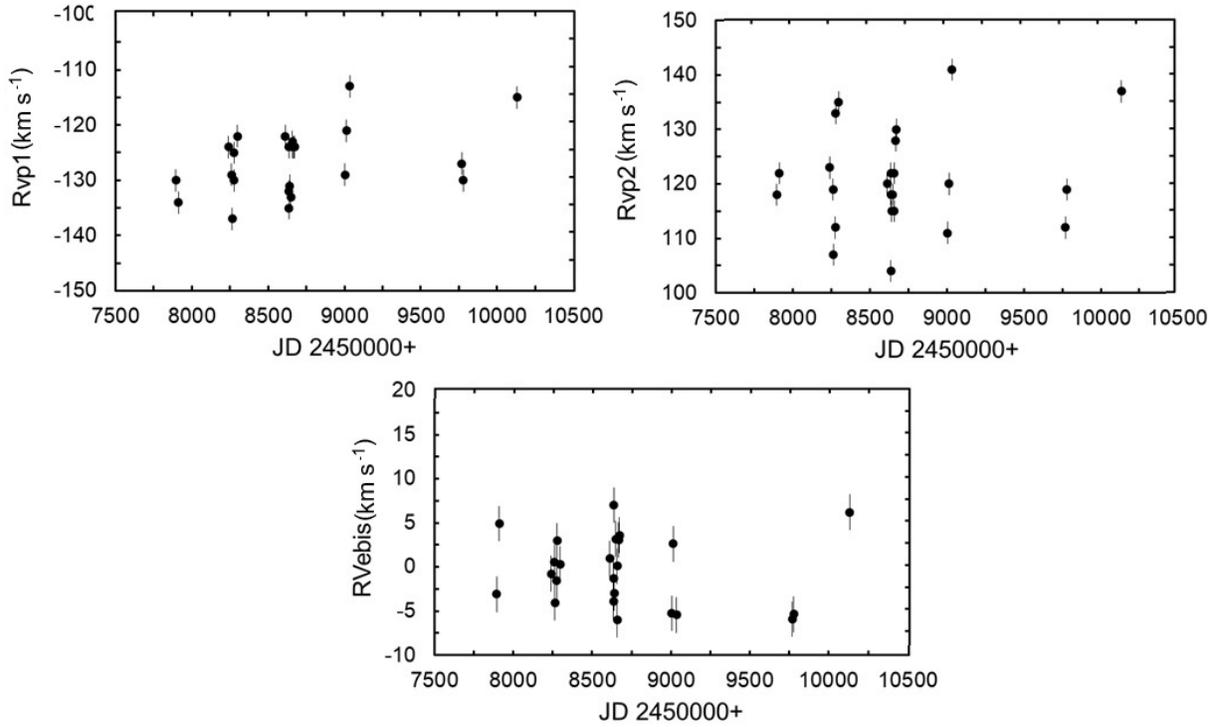

**Figure 4.** Change in RVs of individual components of the Hα emission. Designations of the components are given in Table 3.

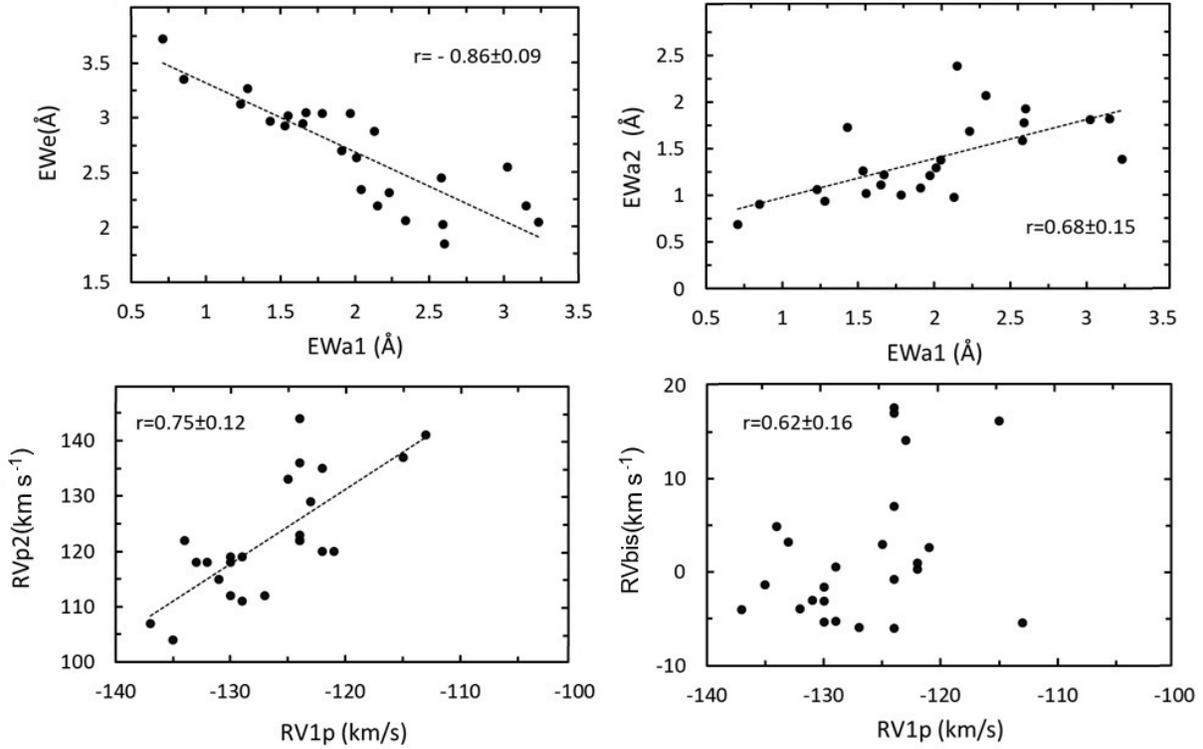

**Figure 5.** Results of the correlation analysis between the parameters of the Hα line. The dotted lines show the approximation by a linear polynomial. The designations of the components are given in Table 3.

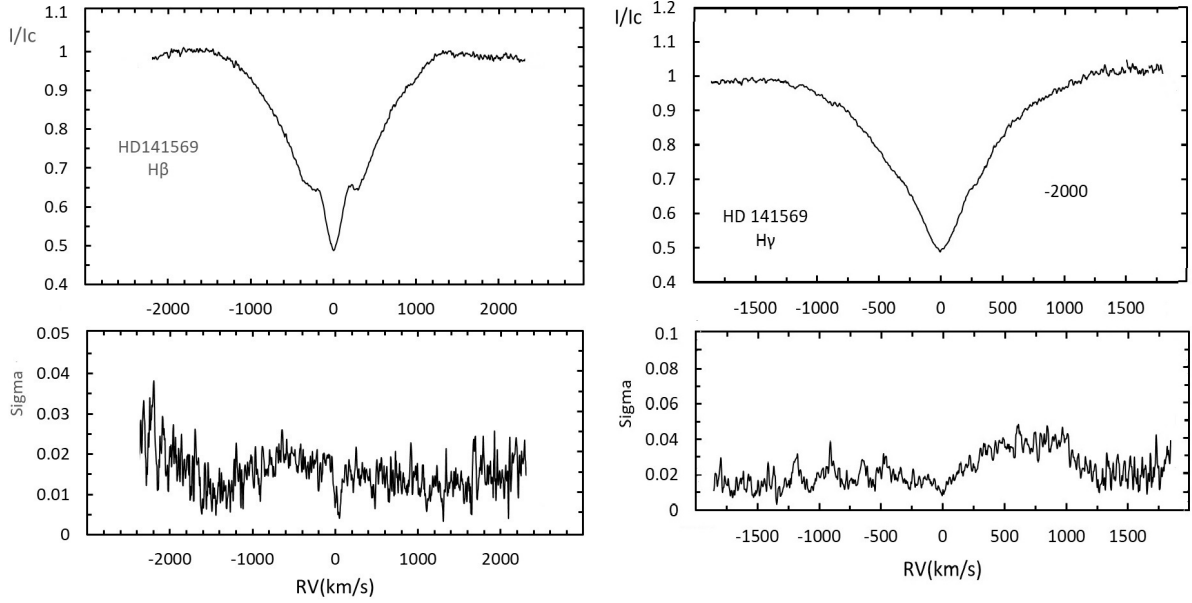

**Figure 6.** Averaged normalized profiles for the Hβ and Hγ lines, and the standard deviation from the average σ by wavelength.

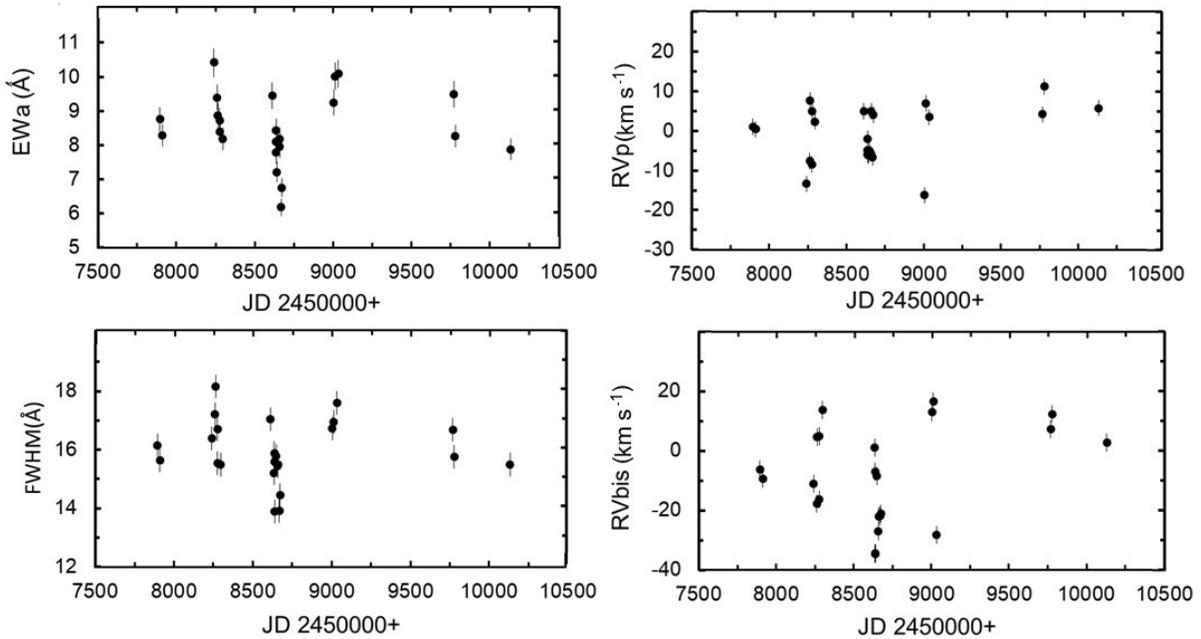

**Figure 7.** For the Hβ line, the dependence of the EW EWa, half-width FWHM, RV of the line peak and bisector velocity RVp1 and RVbis.

Figure 7 depicts the change in the Hβ line parameters over time. As can be seen from here, the line EW shows variability in different years, synchronously with the change in the absorption component of the Hα line. The average value of the EW differs in different years. Minor variability is also observed in other Hα line parameters. The characteristic time of changes is about a year.

Figure 8 displays graphs of changes in the spectrophotometric parameters of the Hε line. As can be seen, the variability is similar to other absorption lines, but on a much smaller scale. This means that with an increase in the depth of the photosphere, a weakening of the star's activity is observed. A smooth increase in most parameters is observed until 2019, and then their decrease. The total time of the full cycle can be ascertained to be about 6–8 yr.

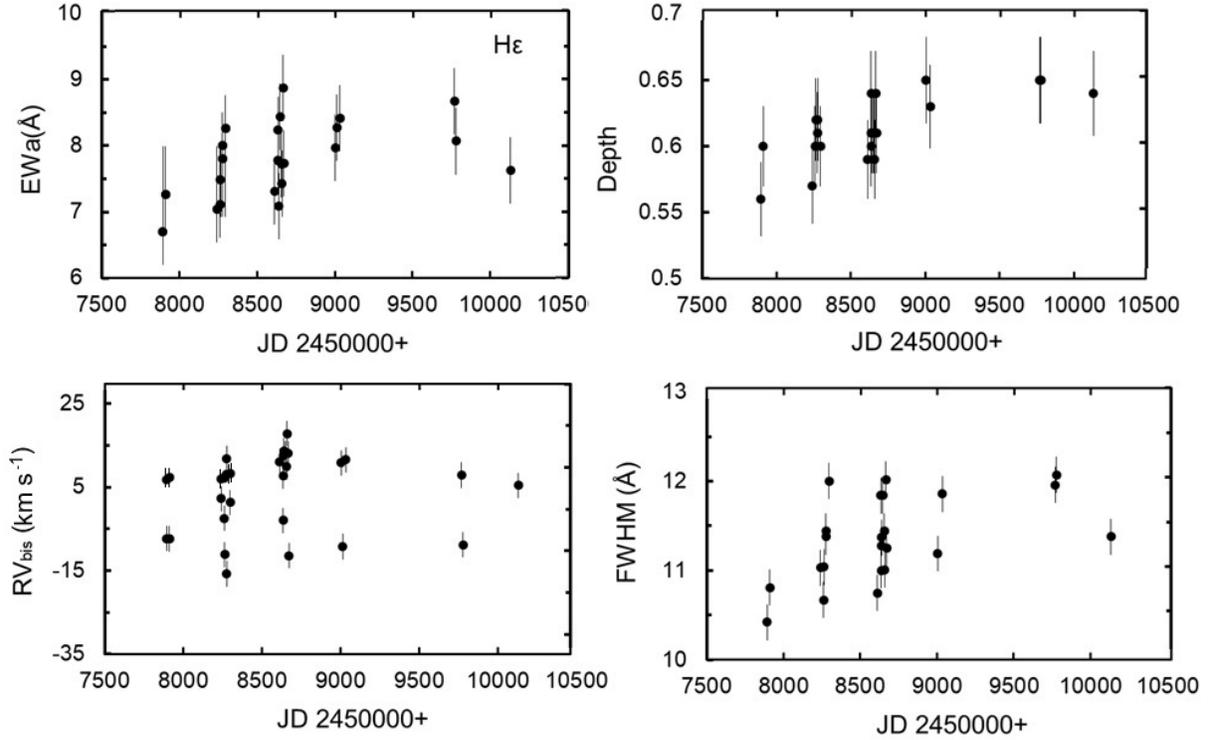

**Figure 8.** Dependence of the EW EWa, line depth (Depth), RV RVbis and FWHM versus time for the H$\varepsilon$ line.

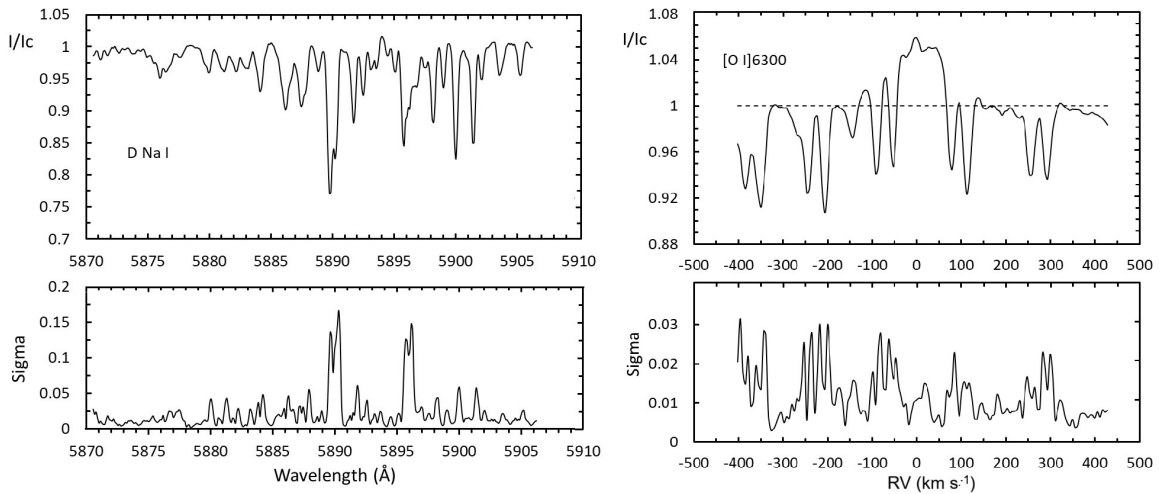

**Figure 9.** Average profiles along the lines of the sodium doublet D Na I and helium He I 5876 (left) and at the [O I] 6300 line (right). As can be seen, the He lines I 5876 and [O I] 6300 showed no noticeable changes.

Figure 9 presents the average profiles of the Na I D doublet lines, He I 5876 Å, and the forbidden line [O I] 6300 Å, compiled from all available data. Based on the standard deviation ($\sigma$) values shown at the bottom of the figure, the Na I $D_1$ and $D_2$ lines exhibit significant variability in their wings. In contrast, no noticeable profile variability was detected in the He I 5876 Å or [O I] 6300 Å lines. According to our measurements, the average FWHM of the [O I] 6300 emission line is approximately $130 \pm 15\,\text{km s}^{-1}$, and the line center is shifted by RVbis = $2 \pm 1.5\,\text{km s}^{-1}$. For comparison,

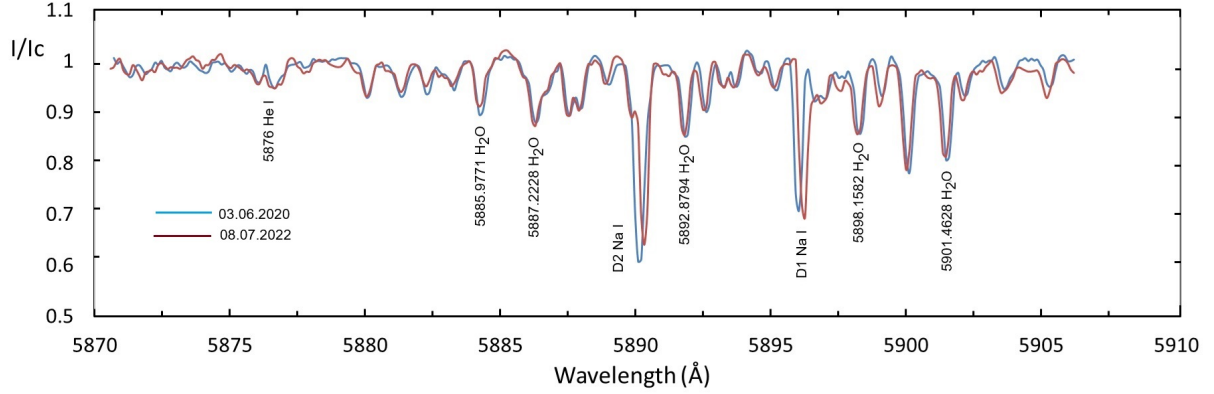

**Figure 10.** Change of position of lines D Na I for 2020 (blue) and 2022 (red).

**Table 4**
Spectrophotometric Parameters Averaged Over All Data for the Spectrum of the Star HD 141569A

| Lines | EWa (Å) | ±σ (Å) | Depth | ±σ | FWHM (Å) | ±σ (Å) | RVp (km s$^{-1}$) | ±σ (km s$^{-1}$) | Rvbis (km s$^{-1}$) | ±σ (km s$^{-1}$) |
|---|---|---|---|---|---|---|---|---|---|---|
| H$\beta$ | 8.51 | 1.02 | 0.51 | 0.03 | 15.97 | 1.06 | −0.81 | 7.07 | −9.88 | 17.57 |
| H$\gamma$ | 7.45 | 0.71 | 0.54 | 0.03 | 12.24 | 0.72 | −8.67 | 12.45 | −2.41 | 13.86 |
| H$\delta$ | 7.84 | 1.09 | 0.59 | 0.03 | 12.17 | 1.27 | −4.77 | 10.28 | −5.95 | 13.72 |
| H$\varepsilon$ | 7.80 | 0.55 | 0.62 | 0.04 | 11.93 | 2.88 | −1.61 | 11.86 | 3.05 | 10.08 |
| He I 5876 | 0.15 | 0.21 | 0.05 | 0.02 | 1.82 | 1.13 | 8.12 | 18.72 | 11.78 | 1 5.62 |
| D2 NaI | 0.41 | 0.27 | 0.33 | 0.09 | 0.47 | 0.22 | −8.98 | 8.62 | −8.79 | 8.73 |
| D1 NaI | 0.26 | 0.10 | 0.26 | 0.08 | 0.39 | 0.10 | −7.92 | 9.25 | −7.75 | 9.14 |

Acke et al. (2005) reported corresponding values of 154 km s$^{-1}$ for the FWHM and a line shift of 10 km s$^{-1}$. Figure 10 shows the star's spectrum in the region of the Na I D doublet and He I 5876 Å lines for two observation dates: 2020 June 3 and 2022 July 8.

Note that according to our data, the [O I] 6300 Å line shows a broad profile with an FWHM of about 200 km s$^{-1}$ and a bisector velocity of about $+5 \pm 3$ km s$^{-1}$. This is consistent with the RV of the star itself, which is obtained from –7.5 to +3 km s$^{-1}$ (see, for example, Matthias et al. 2020; Gaia Collaboration 2022). On certain dates, our observations revealed a profile with two asymmetric peaks, with velocities of –25 km s$^{-1}$ and +50 km s$^{-1}$ for the blue and red peaks respectively. A similar profile for this line was observed in the paper by Acke et al. (2005). In the same paper, it was shown that the [O I] 6300 Å line is formed in the outer part of the circumstellar disk. This line may be formed in previously known asymmetric gas rings in the outer part of the circumstellar disk of HD 141569A (see White et al. 2016).

Telluric H$_2$O lines are visible on blue and red sides of the Na I lines, serving as reference points for checking the spectral line shifts. As seen in the figure, the telluric lines remain stationary, while the peaks of the Na I D lines show a measurable shift of $0.23 \pm 0.01$ Å. The corresponding RVs at the Na I D line peaks are $4 \pm 1$ km s$^{-1}$ in 2020 and $17 \pm 1$ km s$^{-1}$ in 2022. Meanwhile, the structure of the He I 5876 line remains essentially unchanged between these two epochs. This confirms that the observed RV variations over time are specific to the sodium lines.

Our data showed that the variability of the spectroscopic parameters for individual lines occurs with a characteristic time of more than a year, and the period of axial rotation of the star, as we showed in Section 5, is approximately 12.6 hr. The observed variability of the D Na I lines, according to our data, cannot be explained by the axial rotation of the star. Moreover, as can be seen in Table 4, the root-mean-square scatter of the average RV, for all spectra of the sodium doublet lines, is approximately $9.0 \pm 1.5$ km s$^{-1}$, which has approximately the same RV as that of the emission peaks in the H$\alpha$ line (Table 3).

In the classic paper by Finkenzeller & Mundt (1984), it was shown that the D Na I doublet lines in HAeBe stars are indicators of the motion of stellar matter in the circumstellar disk. This suggests that the main cause of the variations in these lines, as in the hydrogen emission lines, is activity in the circumstellar disk of the star.

The D Na I absorption lines may have a significant interstellar contribution due to clouds in the line of sight of HD 141569A (see, e.g., Redfield 2007). Therefore, the EW values given in Table 4 should be considered as an upper limit of the absorption lines. However, the timescale of interstellar extinction variability is significantly larger than the time span covered by our spectra (see, for example, Lauroesch & Meyer 2003). Thus, the observed variability of D Na I lines is most likely due to the gas component in the disk.

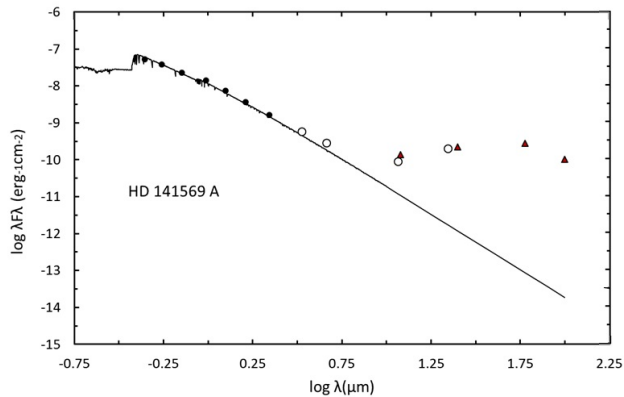

**Figure 11.** The SED curve of the star HD 141569A in the range $\lambda = 0.36$–$100$ $\mu$m. The solid line represents the model SED curve from Castelli & Kurucz (2004). Black circles signify data from 2MASS, open circles are from WISE, and triangles represent data from IRAS.

Table 4 presents the mean values and their standard deviations ($\pm\sigma$) of the spectrophotometric parameters derived for various spectral lines of the star HD 141569A. The table's columns, from left to right, specify: the EW of absorption (EWa), line depth (Depth), FWHM, RV at the peak (RVp), and bisector RV (RVbis) for each line. Over the six years of observations, the relative variation in EWs of hydrogen absorption lines is 10%–12%, which is roughly twice the associated measurement uncertainty. The most substantial variations appear in the helium line He I 5876 Å–up to 140%– and in the sodium doublet: 66% for the $D_2$ line and 38% for the $D_1$ line. RV parameters show even greater variability–up to ten times the magnitude of their measurement errors.

## 5. Physical Parameters of HD 141569A

In the recent work by Adigozalzade et al. (2025a), a spectral energy distribution (SED) curve was constructed and several physical parameters of the star were determined. Figure 11 shows the SED of the star, which we have constructed over the wavelength range of 0.36–100 $\mu$m. The solid curve corresponds to the model of Castelli & Kurucz (2004) with parameters close to the studied object with the solar chemical composition. As shown in Figure 11, IR excess radiation in the SED starts at wavelengths $\lambda \geqslant 10$ $\mu$m. The shape of the energy distribution curve corresponds to a Type II disk according to the classification by Lada & Lada (1989). This suggests the presence of a substantial gas component in the circumstellar disk of the star.

We obtained the following parameters: absolute bolometric luminosity Mvb = $1.2 \pm 0.2$, luminosity $L/L_\odot = 27.7 \pm 0.5$, radius $R/R_\odot = 2.9 \pm 0.3$, mass $M/M_\odot = 2.4 \pm 0.2$, and age $t = 3.5 \pm 0.2$ Myr. Comparison with Table 1 shows that these values agree well with previously published data.

Based on the homogeneous spectral observations of HD 141569A carried out between 2017 and 2023, the following features of the star's spectral variability have been identified:

1. Long-term smooth variability in the hydrogen absorption lines was detected, with a characteristic timescale of approximately 6–8 yr.
2. Relatively rapid, short-term variability was observed, which occurring over several months.
3. No fast variability on daily or weekly timescales was detected.

Changes in their FWHM accompany the long-term variability observed in the EWs of hydrogen absorption lines, but not by significant changes in the RVs of these lines. Furthermore, both the peak RVs and the bisector velocities of the hydrogen lines show only minor fluctuations, with amplitudes around $\pm 10$ km s$^{-1}$.

To explain these variations, we propose that the circumstellar disk has a complex, inhomogeneous structure, which often partially obscures the photospheric radiation of the central star. These inhomogeneous clumps, likely composed of gas and dust, appear to be large and extended, capable of partially blocking stellar radiation for periods of three years or more. They are likely semi-transparent at the edges, causing smooth transitions into and out of partial eclipses. Such clumps may represent fragmented regions of the circumstellar disk.

These passing gas clumps can also generate additional emission in the hydrogen lines, altering the line profiles and affecting the bisector RV, which is measured relative to the wings of the profile at half intensity. The appearance of this additional emission is evidenced by changes in spectra:

1. The line EWs
2. The depth and FWHM of the H$\alpha$ and other hydrogen lines
3. The wing asymmetry of the average hydrogen line profiles
4. The FWHMs of absorption lines

All of these features have been confirmed in our spectral measurements.

The profile of the [O I] 6300 Å forbidden line remained virtually unchanged throughout the observational period, indicating the presence of a stationary gas structure around the star. Such structures are typical for $\beta$ Pic and Vega-type stars (see, e.g., Janson et al. 2021).

Based on the derived physical parameters and stellar age, HD 141569A appears to be in a transitional evolutionary stage, evolving from a protoplanetary disk to a dissipating circumstellar disk. The disk currently contains relatively little gas, while excess emission from cold dust dominates in the far-IR range. The presence of residual gas in the disk was also confirmed by Zuckerman et al. (1995), Thi et al. (2014), and Di Folco et al. (2020).

Disk formation around young stars is a generally accepted fact, driven by the fact that after the formation of a young protostar from primordial molecular clouds, according to the laws of celestial mechanics, a disk should form around the star to preserve its primordial moment of inertia (see, for example, Concha-Ramírez et al. 2023). Since HD 141569A is a rapidly rotating star, it was interesting to test the ratio of the equatorial and critical velocities $V_e$ and $V_c$ to determine whether the star is losing matter due to rapid rotation. Observations of the gas disk structure indicate an inclination angle of $57° \pm 1°$ relative to the line of sight (Di Folco et al. 2020). From our data, the stellar radius is estimated as $R_* = 2.9 \pm 0.3\ R_\odot$. If we adopt the projected rotational velocity $v\sin i = 236 \pm 15$ km s$^{-1}$ (Merín et al. 2004), then the calculated equatorial rotational velocity is $V_e = 281 \pm 15$ km s$^{-1}$. From here, we can calculate the period of axial rotation

$$P = \frac{2\pi R_*}{V_e}. \quad (1)$$

The star's period of rotation is estimated to be $P = 45367$ second, or approximately 12.6 hr. Compared to other studies, our derived stellar radius appears to be somewhat overestimated. For example, if we adopt a smaller radius of 1.5 $R_\odot$ as reported by Fairlamb et al. (2015), the corresponding rotation period becomes $P \approx 9.3$ hr. This implies that using a smaller radius leads to a shorter rotation period and, consequently, a lower equatorial rotation velocity. The critical rotation velocity (i.e., the velocity at which the centrifugal force balances gravity at the equator) can be estimated using the following relation

$$V_c = \sqrt{\frac{GM_*}{R_e}}, \quad (2)$$

where $G = 6.67 \cdot 10^{-11}$ m$^3$ s$^{-2}$ kg$^{-1}$, $M = 2.9\ M_\odot$, and $R_e = 2.9\ R_\odot$. The critical rotational velocity of the star is estimated to be $V_c = 436 \pm 15$ km s$^{-1}$, which is significantly higher than the star's equatorial rotational velocity $V_e$. This suggests that the matter ejected by the stellar wind could not reach the escape velocity and should mostly fall back onto the stellar surface because there is a nonstationary balance between centrifugal force and gravity.

## Acknowledgments

The authors thank the reviewers for their helpful comments, which contributed significantly to improving the article.

## ORCID iDs

N. Z. Ismailov  https://orcid.org/0000-0002-5307-4295